\documentclass[twocolumn,aps,prd,showpacs,10pt,superscriptaddress]{revtex4-1}
\usepackage{amssymb}
\usepackage{amsmath}
\newcommand{\vect}[1]{\boldsymbol{#1}} 
\usepackage{multirow}
\usepackage{graphicx}
\usepackage[usenames,dvipsnames,svgnames]{xcolor}
\usepackage{hyperref}
\hypersetup{
pdfnewwindow=true,      
colorlinks=true,        
linkcolor=Blue,         
citecolor=Blue,         
filecolor=Blue,         
urlcolor=Blue           
}

\begin{document}

\title{Homogeneous nonequilibrium molecular dynamics method for heat transport and spectral decomposition with many-body potentials}
\author{Zheyong Fan}
\email{Corresponding author: brucenju@gmail.com} 
\affiliation{School of Mathematics and Physics, Bohai University, Jinzhou, China}
\affiliation{QTF Centre of Excellence, Department of Applied Physics, Aalto University, FI-00076 Aalto, Finland}
\author{Haikuan Dong}
\affiliation{School of Mathematics and Physics, Bohai University, Jinzhou, China}
\author{Ari Harju}
\affiliation{QTF Centre of Excellence, Department of Applied Physics, Aalto University, FI-00076 Aalto, Finland}
\author{Tapio Ala-Nissila}
\affiliation{QTF Centre of Excellence, Department of Applied Physics, Aalto University, FI-00076 Aalto, Finland}
\affiliation{Centre for Interdisciplinary Mathematical Modelling and Department of Mathematical Sciences, Loughborough University, Loughborough, Leicestershire LE11 3TU, UK}

\date{Feb 11, 2019}

\begin{abstract}
The standard equilibrium Green-Kubo and nonequilibrium molecular dynamics (MD) methods for computing thermal transport coefficients in solids typically require relatively long simulation times and large system sizes. To this end, we revisit here
the homogeneous nonequilibrium MD method by Evans [Phys. Lett. A \textbf{91}, 457 (1982)] and generalize it to many-body potentials that are required for more realistic materials modeling.  We also propose
a method for obtaining spectral conductivity and phonon mean free path from the simulation data. This spectral decomposition method does not require lattice dynamics calculations and can find important applications in spatially complex structures.
We benchmark the method by calculating thermal conductivities of three-dimensional silicon, two-dimensional graphene, and a quasi-one-dimensional carbon nanotube and show that
the method is about one to two orders of magnitude more efficient than the Green-Kubo method. We apply the spectral decomposition method to examine the long-standing dispute over thermal conductivity convergence vs. divergence in carbon nanotubes.
\end{abstract}

\maketitle

\section{Introduction}
Heat transport at the nanoscale \cite{cahill2014apr,volz2016epjp} is vital for many technological applications such as thermal management of electronic devices, thermoelectric energy conversion, and nanoparticle-mediated thermal therapy, just to name a few.
Molecular dynamics (MD) is the most complete classical method to study heat transport at the nanoscale. All the MD based methods for computing the heat transport coefficient, namely, the thermal conductivity, are fundamentally based on Fourier's law $Q^{\mu}=-\sum_{\nu}\kappa^{\mu\nu}\partial T /\partial x_{\nu}$, where $Q^{\mu}$ is the heat flux in the $\mu$ direction, $\partial T /\partial x_{\nu}$ is the temperature gradient in the $\nu$ direction, and $\kappa^{\mu\nu}$ is the $\mu\nu$ component of the thermal conductivity tensor.
Methods directly based on this are called nonequilibrium MD (NEMD) methods and have a few variants  \cite{hoover1975tcap,ikeshoji1994mp,jund1999prb,muller1997jcp}. When the purpose is to compute the length-convergent thermal conductivity in the diffusive regime, the NEMD methods are computationally inefficient for good thermal conductors \cite{liang2015jap}, because one needs to compute the thermal conductivities of several systems with lengths exceeding the effective phonon mean free path for an accurate extrapolation \cite{sellan2010prb}. The approach-to-equilibrium MD (AEMD) method proposed recently \cite{lampin2013jap,melis2014epjb} has a similar disadvantage \cite{zaoui2016prb}. In both the NEMD and the AEMD methods, the phonon transport is affected by boundary scattering due to the inhomogeneity introduced by the high and low temperature regions.

There also exist homogeneous MD methods where boundary scattering in the transport direction is absent. The equilibrium MD (EMD) method \cite{levesque1973pra} based on the Green-Kubo relation \cite{green1954jcp,kubo1957jpsj,mcquarrie2000book} derived from linear response theory is the most popular one. Due to the absence of boundary scattering, one only needs to use a simulation cell that is large enough to accommodate the major phonon wavelengths \cite{wang2017jap}. In the EMD method, the thermal conductivity is calculated as an integral of the heat current autocorrelation function. It is well known that accurate evaluation of time correlation functions in MD is computationally demanding, due to the increasing noise-to-signal ratio with increasing correlation time. 

In 1982 Evans \cite{evans1982pla} proposed a different approach called the homogeneous nonequilibrium MD (HNEMD) method. It is a nonequilibrium method because external forces are added to the system. It is also a homogeneous method because no temperature gradient is generated. Evans' original method was derived for two-body potentials only. More recently, it was used with the Tersoff \cite{tersoff1989prb} and Brenner potentials \cite{brenner2000jpcm} to calculate the thermal conductivity of carbon nanotubes \cite{berber2000prl,lukes2006jht}. However, these works do not derive an extension of the method to many-body potentials. This is in fact a nontrivial matter as demonstrated by Mandadapu \textit{et al.} \cite{mandadapu2009jcp,mandadapu2009pre,mandadapu2010jcp}. Most importantly, Refs. \cite{berber2000prl,lukes2006jht} used unphysically large external forces such that the linear response theory itself is no longer valid as we will explicitly demonstrate here. 

The main purpose of the present paper is to rigorously derive the HNEMD method for systems described by many-body empirical potentials, to discuss various technical issues on the proper use of this method in practice, and to propose a novel spectral decomposition method for obtaining the spectral conductivity $\kappa(\omega)$ and the phonon mean free path $\lambda(\omega)$ in the diffusive regime. The spectral decomposition method does not need lattice dynamics calculations in contrast to the existing ones \cite{ladd1986prb,mcgaughey2004prb,henry2008jctn,turney2009prb_bte,thomas2010prb,feng2015jap,gill-comeau2015prb,lv2016njp} based on the EMD method. We note that the generalization of the HNEMD method from two-body to many-body potentials has been previously considered by Mandadapu \textit{et al.} \cite{mandadapu2009jcp,mandadapu2009pre,mandadapu2010jcp}, but their formalism only applies to a special class of many-body potentials called cluster potentials \cite{tadmor2011book,martin1975jpc}, including the Stillinger-Weber \cite{stillinger1985prb} potential, but not to more general many-body potentials such as the Tersoff potential. In the present work we present a general derivation valid for all many-body potentials. We then benchmark the method for various model systems and demonstrate that it is about two orders of magnitude more efficient than the EMD one.

\section{HNEMD method for general many-body potentials}

\subsection{Derivations based on linear response theory}

We first derive the thermal conductivity expression in the HNEMD method. Consider a system of $N$ particles described by the general Hamiltonian
\begin{equation}
\label{equation:H}
H(\{\vect{r}_i,\vect{p}_i\}) = \sum_{i} \frac{\vect{p}_i^2}{2m_i} + U(\{\vect{r}_i\}),
\end{equation}
with the equations of motion 
$d\vect{r}_i/dt = \vect{p}_i/m_i$ and $d\vect{p}_i/dt = \vect{F}_i$. Here, $\vect{r}_i$, $m_i$, and $\vect{p}_i$ are the position, mass, and momentum of particle $i$, and $\vect{F}_i$ is the total force acting on it. In the linear response theory \cite{evans1990book,tuckerman2010book}, one introduces a driving force and the equations of motion are modified to
\begin{equation}
\label{equation:eom-r}
\frac{d\vect{r}_i}{dt} = \frac{\vect{p}_i}{m_i} + \mathbf{C}_i(\{\vect{r}_i,\vect{p}_i\}) \cdot \vect{F}_{\rm e};
\end{equation}
\begin{equation}
\label{equation:eom-p}
\frac{d\vect{p}_i}{dt} = \vect{F}_i + \mathbf{D}_i(\{\vect{r}_i,\vect{p}_i\}) \cdot \vect{F}_{\rm e}.
\end{equation}
Here $\mathbf{C}_i(\{\vect{r}_i,\vect{p}_i\})$ and $\mathbf{D}_i(\{\vect{r}_i,\vect{p}_i\})$ are tensors of rank two and $\vect{F}_{\rm e}$ is a vector. The total time derivative of the Hamiltonian can be written as \cite{evans1990book,tuckerman2010book}
\begin{equation}
\label{equation:dHdt}
\frac{dH(\{\vect{r}_i,\vect{p}_i\})}{dt} = \vect{J}_{\rm d} \cdot \vect{F}_{\rm e},
\end{equation}
where $\vect{J}_{\rm d}=\vect{J}_{\rm d}(\{\vect{r}_i,\vect{p}_i\})$ is called the dissipative flux vector. In terms of the dissipative flux, the nonequilibrium ensemble average $\langle \rangle_{\rm ne}$ of a general vector physical quantity $\vect{A}(\{\vect{r}_i,\vect{p}_i\})$ at time $t$ after switching on the external driving force can be written as \cite{evans1990book,tuckerman2010book} ($k_{\rm B}T$ is the thermal energy)
\begin{align}
\langle \vect{A}(t)\rangle_{\rm ne}=\langle \vect{A}(0)\rangle 
+\left( \int_0^tdt'\frac{\langle \vect{A}(t')\otimes \vect{J}_{\rm d}(0)\rangle}{k_{\rm B}T} \right)
\cdot \vect{F}_{\rm e}.
\label{equation:A(t)}
\end{align}
Here, $\langle \vect{A}(0)\rangle$ is the usual equilibrium ensemble average of $\vect{A}$ and $\langle \vect{A}(t')\otimes \vect{J}_{\rm d}(0)\rangle$ is the equilibrium time correlation function between $\vect{A}$ and $\vect{J}_{\rm d}$.

The central idea of the HNEMD method by Evans \cite{evans1982pla} is to set both $\vect{A}$ and $\vect{J}_{\rm d}$ in Eq. (\ref{equation:A(t)}) to the heat current operator $\vect{J}_{\rm q}$, giving (note that $\langle \vect{J}_{\rm q}(0)\rangle=0$)
\begin{equation}
\langle \vect{J}_{\rm q}(t)\rangle_{\rm ne}
=
\left(
\frac{1}{k_{\rm B}T}\int_0^tdt'\langle \vect{J}_{\rm q}(t')\otimes \vect{J}_{\rm q}(0)\rangle
\right)
\cdot \vect{F}_{\rm e}.
\label{equation:J(t)}
\end{equation}
where $\langle \vect{J}_{\rm q}(t')\otimes \vect{J}_{\rm q}(0)\rangle$ is the equilibrium heat current autocorrelation function. Setting $\vect{J}_{\rm d}$ to $\vect{J}_{\rm q}$ fixes the equations of motion, as we will discuss soon. According to the Green-Kubo relation \cite{green1954jcp,kubo1957jpsj,mcquarrie2000book}, the quantity in the parentheses is related to the (running) thermal conductivity tensor, 
\begin{equation}
   \kappa^{\mu\nu}(t) = \frac{1}{k_{\rm B}T^2V} \int_0^t dt'\langle J_{\rm q}^{\mu}(t') J_{\rm q}^{\nu}(0)\rangle,
\end{equation}
$V$ being the system volume. Therefore, Eq. (\ref{equation:J(t)}) can be interpreted as 
\begin{equation}
\frac{\langle J^{\mu}_{\rm q}(t)\rangle_{\rm ne}}{TV} = \sum_{\nu} \kappa^{\mu\nu}(t) F_{\rm e}^{\nu}.    
\end{equation}
Working with principal axes \cite{nye1957book}, the thermal conductivity tensor is diagonal and the thermal conductivity $\kappa$ in a given direction is given by 
\begin{equation}
\label{equation:kappa}
\kappa(t) = \frac{\langle J_{\rm q}(t)\rangle_{\rm ne}}{TV F_{\rm e}}.    
\end{equation}
The running thermal conductivity $\kappa(t)$ calculated using this equation will show large fluctuations and it is not easy to judge when $\kappa(t)$ has converged. One can circumvent this difficulty by redefining $\kappa(t)$ as the following cumulative average:
\begin{equation}
\label{equation:kappa_prime}
\kappa(t)= \frac{1}{t} \int_0^{t} ds \frac{\langle J_{\rm q}(s)\rangle_{\rm ne}}{TV F_{\rm e}}.
\end{equation}
A similar definition has been implicitly used in previous works \cite{mandadapu2009jcp,dongre2017msmse} on the HNEMD method. 

To complete the derivation of the generalized HNEMD method, we need to determine the equations of motion, which are the foundation of the MD simulations. They are closely related to the heat current $\vect{J}_{\rm q}$ when the dissipative flux $\vect{J}_{\rm d}$ defined in Eq. (\ref{equation:dHdt}) is chosen to be the same as $\vect{J}_{\rm q}$. We discuss the heat current and the equations of motion next.

The general heat current formulae in MD simulations have been discussed in Ref. \cite{fan2015prb} in great detail. For a general many-body potential with the total potential energy $U=\sum_i U_i(\{\vect{r}_{ij}\}_{j\neq i})$, the heat current can be written as \cite{fan2015prb}
\begin{equation}
\label{equation:J_many-body}
\vect{J}_{\rm q} = \vect{J}_{\rm q}^{\rm kin}  + \vect{J}_{\rm q}^{\rm pot} 
= \sum_i \frac{\vect{p}_i}{m_i} E_i
+ \sum_{i,j\neq i} \frac{\vect{p}_i}{m_i} \cdot \left(\frac{\partial U_j}{\partial \vect{r}_{ji}} \otimes \vect{r}_{ij} \right),
\end{equation}
where $E_i=\vect{p}_i^2/2m_i+U_i$ is the total energy of particle $i$ and $U_i$ is the potential energy. The position difference is defined as $\vect{r}_{ij} \equiv \vect{r}_j - \vect{r}_i$. The equations of motion are constructed to make the dissipative flux $\vect{J}_{\rm d}$ identical to the heat current $\vect{J}_{\rm q}$. Evans chose the term $\mathbf{C}_i(\{\vect{r}_i,\vect{p}_i\})=0$. Then, the time derivative of the Hamiltonian (\ref{equation:H}) can be derived from the equations of motion (\ref{equation:eom-r}) and (\ref{equation:eom-p}) to be
\begin{equation}
\frac{dH}{dt} =\sum_i \frac{\vect{p}_i}{m_i} \cdot \left(\mathbf{D}_i\cdot \vect{F}_{\rm e} \right).    
\end{equation}
Comparing this with Eqs. (\ref{equation:dHdt}) and (\ref{equation:J_many-body}) and setting $\vect{J}_{\rm d}=\vect{J}_{\rm q}$, we have
\begin{equation}
\label{equation:driving_force}
\mathbf{D}_i \cdot \vect{F}_{\rm e} = E_i \vect{F}_{\rm e} +  \sum_{j \neq i} \left(\frac{\partial U_j}{\partial \vect{r}_{ji}} \otimes \vect{r}_{ij}\right) \cdot \vect{F}_{\rm e}.    
\end{equation}
This driving force will be added to the total force for particle $i$. Because the summation $\sum_i \mathbf{D}_i \cdot \vect{F}_{\rm e} \neq 0$, the total momentum of the system will not be conserved under this driving force. To restore momentum conservation, one needs to subtract the mean force of the total system from the force on each particle. Formally, this is equivalent to modifying the driving force to 
\begin{align}
\label{equation:DF-many-body}
\mathbf{D}_i \cdot \vect{F}_{\rm e}
&= E_i\vect{F}_{\rm e} -\frac{1}{N}\sum_jE_j \vect{F}_{\rm e} \nonumber \\
&+ \sum_{j \neq i} \left(\frac{\partial U_j}{\partial \vect{r}_{ji}} \otimes \vect{r}_{ij}\right) \cdot \vect{F}_{\rm e} \nonumber \\
&- \frac{1}{N}\sum_j \sum_{k \neq j} \left(\frac{\partial U_k}{\partial \vect{r}_{kj}} \otimes \vect{r}_{jk}\right) \cdot \vect{F}_{\rm e}.
\end{align}
One can easily verify that for two-body potentials, Eq. (\ref{equation:DF-many-body}) reduces to that by Evans \cite{evans1982pla}. However, we emphasize that the heat current formula for two-body potentials does not apply to many-body potentials \cite{fan2015prb}. One also needs to apply a thermostat to keep the temperature of the system at the target. To this end, we use the Nos\'{e}-Hoover chain thermostat \cite{tuckerman2010book} here. 

\subsection{Explicit algorithm}

After deriving the formalism of the HNEMD method for thermal conductivity calculations using general many-body potentials, we present an explicit algorithm which can be readily implemented in a computer. 

An HNEMD simulation consists of the following steps:

\begin{enumerate}
\item Equilibration. First, as in any MD simulation, we equilibrate the system in the $NVT$ or the $NpT$ ensemble to reach thermal equilibrium. Note that as in the EMD method, periodic boundary conditions must be applied to the transport direction. 
\item Production. Second, we generate the homogeneous heat current by adding a driving force as given by Eq. (\ref{equation:driving_force})
on top of the interatomic force \cite{fan2015prb}
\begin{equation}
\vect{F}_i = \sum_{j\neq i} \left(\frac{\partial U_i}{\partial \vect{r}_{ij}} - \frac{\partial U_j}{\partial \vect{r}_{ji}} \right),
\end{equation}
to get the total force
\begin{align}
\vect{F}_{i}^{\rm tot}
= \vect{F}_i + \mathbf{D}_i \cdot \vect{F}_{\rm e}.
\end{align}
One has to subtract the mean force of the total system from the force on each particle such that the total momentum of the system is conserved. Specifically, we make the correction:
\begin{equation}
\vect{F}_i^{\rm tot} \rightarrow \vect{F}_i^{\rm tot} - \frac{1}{N} \sum_i \vect{F}_i^{\rm tot}.
\end{equation}
At this stage, one also needs to apply a thermostat to keep the temperature of the system at the target; otherwise the system will be heated up by the driving force.
\item Post-processing. Finally, we sample the heat current as given by Eq. (\ref{equation:J_many-body}) and calculate the thermal conductivity according to Eq. (\ref{equation:kappa_prime}). 
\end{enumerate}

\section{Validation and benchmark}

\subsection{Details on the MD simulations}

\begin{figure*}[hbt]
\begin{center}
\includegraphics[width=2\columnwidth]{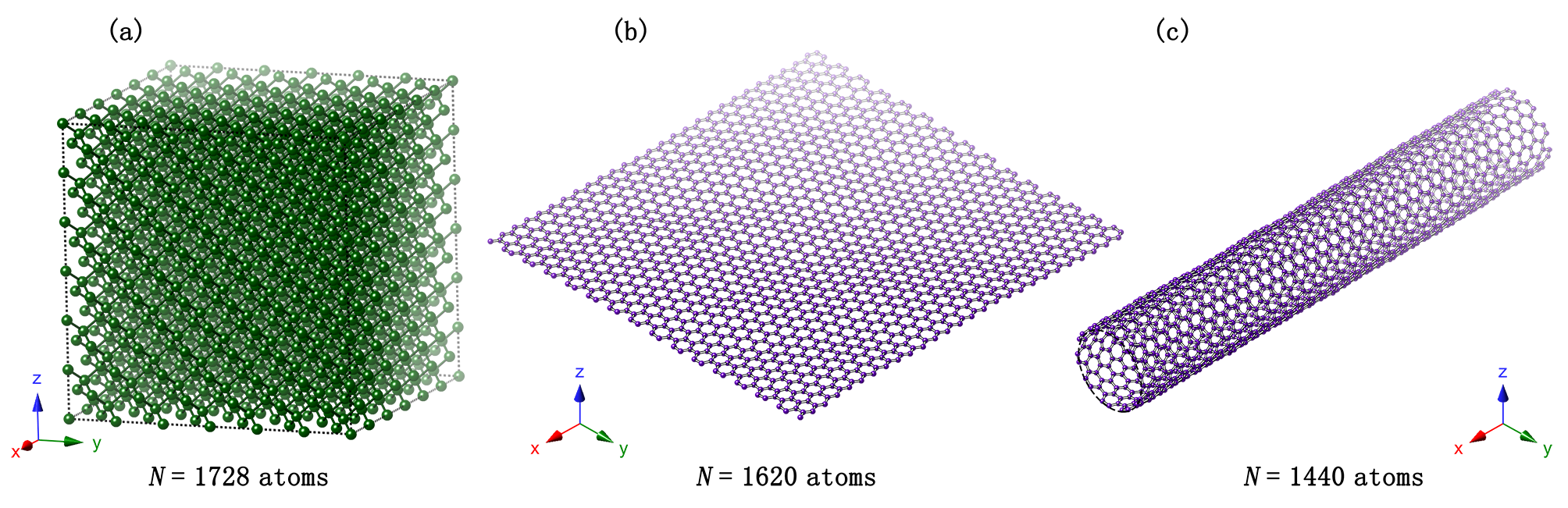}
\caption{Schematic  illustration  of  the  model  systems  studied  in  this  work:  (a)  3D  bulk  silicon;  (b)  2D  graphene;  (c)  Q1D $(10,10)$-CNT. The cell size shown here for silicon is the same as that used in the MD simulations, but for clarity, the cell sizes for graphene and CNT shown here are smaller than those used in the MD simulations.}
\label{figure:structure}
\end{center}
\end{figure*}

The HNEMD method as described above has been implemented in the open source GPUMD (Graphics Processing Units Molecular Dynamics) package \cite{fan2017cpc,fan2017gpumd}. We use it to benchmark the HNEMD method by computing the thermal conductivities of three materials at 300 K and zero pressure: three-dimensional (3D) silicon, two-dimensional (2D) graphene, and a quasi-one-dimensional (Q1D) $(10,10)$-CNT (carbon nanotube). The system in the HNEMD method is in a homogeneous nonequilibrium state because there is no explicit heat source and sink and heat flows circularly under the driving force. Because of the absence of heat source and sink, no boundary scattering occurs for the phonons and the HNEMD method is similar to the EMD method in terms of finite-size effects. Usually, a relatively small simulation cell is thus enough to eliminate them. We use a cubic simulation cell with 1728 atoms for silicon, an almost square-shaped cell with 24 000 atoms for graphene, and a cell with 16 000 atoms for the $(10, 10)$-CNT, all of which are sufficiently large. See Fig. \ref{figure:structure} for an illustration of the atomic structures and  lattice orientations in these model systems.
Periodic boundary conditions are applied in all  the  directions for silicon, the planar directions (the $xy$ plane) for graphene and the axial direction (the $x$ direction) for CNT. For all the systems, the velocity-Verlet integration scheme \cite{tuckerman2010book} with a time step of 1 fs is used. We first equilibrate each system for 2 ns and then apply the external force for 20 ns. The Tersoff potential with parameters from Ref. \cite{tersoff1989prb} is used for silicon and the Tersoff potential with parameters from Ref. \cite{lindsay2010prb} is used for graphene and CNT. An effective thickness of $0.335$ nm for the atom layer in graphene and CNT is used in calculating the volume in these systems. 

\subsection{Cumulative average of the running thermal conductivity \label{section:running}}

\begin{figure}[hbt]
\begin{center}
\includegraphics[width=\columnwidth]{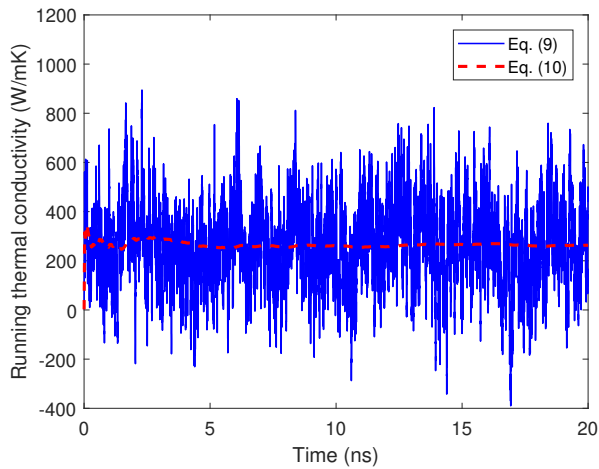}
\caption{Running thermal conductivity as defined in Eq. (\ref{equation:kappa}) and its cumulative average as defined in Eq. (\ref{equation:kappa_prime}) as a function of time $t$ in the nonequilibrium production stage of the MD simulation. The system is a silicon crystal at 300 K and the driving force parameter is $F_{\rm e}=0.3$ $\mu$m$^{-1}$.}
\label{figure:running}
\end{center}
\end{figure}

The running thermal conductivity $\kappa(t)$ calculated using Eq. (\ref{equation:kappa}) for silicon with $F_{\rm e}=0.3$ $\mu$m$^{-1}$ is shown as the solid line (with large fluctuations) in Fig. \ref{figure:running}. Because of the large fluctuations, it is not easy to determine when $\kappa(t)$ has converged. To circumvent this, we redefine $\kappa(t)$ as the cumulative average of the running thermal conductivity, as given by Eq. (\ref{equation:kappa_prime}). The cumulative average of the running thermal conductivity is shown as the dashed line in Fig. \ref{figure:running} and it converges well in the long time limit. This simply means that the ensemble average $\langle \rangle_{\rm ne}$ can be represented as a time average in the MD simulation.

\begin{figure}[hbt]
\begin{center}
\includegraphics[width=\columnwidth]{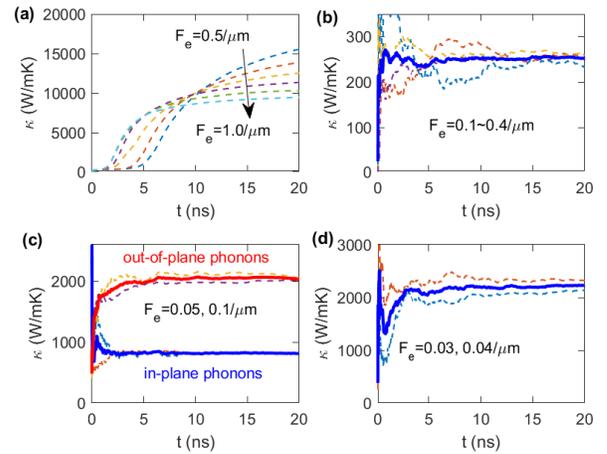}
\caption{Running average $\kappa(t)$ of the thermal conductivity as defined in Eq. (\ref{equation:kappa_prime}) of bulk silicon (a and b), graphene (c), and (10,10)-CNT (d) at 300 K as a function of time $t$. In each subplot, the dashed lines are from individual runs with a given $F_{\rm e}$ and the solid line is the average of them. }
\label{figure:hnemd_si_gra_cnt}
\end{center}
\end{figure}

\subsection{Choice of driving force}

It is known from previous works \cite{evans1982pla,mandadapu2009jcp,dongre2017msmse} that the parameter $F_{\rm e}$ (of dimension inverse length) is crucial: it has to be small enough to keep the system within the linear response regime, and large enough to retain a sufficiently large signal-to-noise ratio. Mandadapu \textit{et al.} \cite{mandadapu2009jcp} have given a rule-of-thumb to determine appropriate values of $F_{\rm e}$: it should be much smaller than $1/\lambda$, where $\lambda$ can be regarded as a characteristic phonon mean free path (MFP) of the system. From our spectral decomposition results (see below), linear response is completely assured when $F_{\rm e} \lambda_{\rm max} \lesssim 1$, where $\lambda_{\rm max}$ is the maximum phonon MFP.

\subsection{Results for silicon}

For silicon crystal  described by the Tersoff potential at 300 K, Figs. \ref{figure:hnemd_si_gra_cnt}(a) and \ref{figure:hnemd_si_gra_cnt}(b) show that $\kappa(t)$ behaves unexpectedly when $F_{\rm e} > 0.4$ $\mu$m$^{-1}$ and converges to reasonable values when $F_{\rm e} \leq 0.4$ $\mu$m$^{-1}$. If we consider a simulation time up to $t=2.5$ ns, which is comparable to the simulation times used in previous works \cite{mandadapu2009jcp,dongre2017msmse}, $\kappa(t=2.5~\rm ns)$ gradually increases with increasing $F_{\rm e}$, similar to the observations in previous works \cite{mandadapu2009jcp,dongre2017msmse}. When considering a long simulation time of $t=20$ ns, $\kappa(t=20~\rm ns)$ first jumps to a very large value at $F_{\rm e} = 0.5$ $\mu$m$^{-1}$ and then decreases with increasing $F_{\rm e}$. The abrupt jump is helpful for quickly identifying the linear response regime. When the system is in the linear response regime, $\kappa(t)$ converges in the long time limit and the converged value does not depend on $F_{\rm e}$ in a systematic way. Using the $\kappa(t=20~\rm ns)$ values with $F_{\rm e} \leq 0.4$ $\mu$m$^{-1}$, the thermal conductivity of silicon at 300 K is determined to be $\kappa=252\pm 7$ W/mK. This is in excellent agreement with the value $\kappa=250\pm 10$ W/mK obtained using the EMD method \cite{dong2018prb}. It should be noted that 50 independent simulations (each with a production time of 20 ns) were used in the EMD calculations \cite{dong2018prb}, while we only need a few simulations in the HNEMD method to achieve comparable accuracy. 

\subsection{Results for graphene}

For graphene [cf. Fig. \ref{figure:hnemd_si_gra_cnt}(c)], we separately calculate \cite{matsubara2017jcp,fan2017prb} the thermal conductivity contributed by the in-plane and out-of-plane (flexural) phonons. The in-plane contribution comes from the terms with $v_x$ and $v_y$ in $\vect{J}^{\rm pot}_{\rm q}$ and the out-of-plane contribution comes from the terms with $v_z$. For details on the thermal conductivity decomposition, see Appendix \ref{section:decomposition}. We have checked that the system is in the linear response regime when $F_{\rm e} \leq 0.1$ $\mu$m$^{-1}$. The converged thermal conductivity is estimated to be $815\pm 23$ W/mK for the in-plane phonons and $2032\pm 26$ W/mK for the out-of-plane phonons. In total, the thermal conductivity of graphene at 300 K is $2847\pm 49$ W/mK, which is in excellent agreement with the EMD value of $2900\pm 100$ W/mK from Ref. \cite{fan2017prb}. The EMD results from Ref. \cite{fan2017prb} were obtained using a total production time of 5000 ns. In contrast, the HNEMD results here were obtained using a total production time of 40 ns, about two orders of magnitude shorter. 

\subsection{Results for $(10,10)$-CNT}

We finally consider the $(10,10)$-CNT [cf. Fig. \ref{figure:hnemd_si_gra_cnt}(d)]. The system is in the linear response regime when $F_{\rm e} \leq 0.04$ $\mu$m$^{-1}$, where $\kappa(t)$ with different $F_{\rm e}$ values converge to comparable values in the long time limit. The converged thermal conductivity is estimated to be $2230\pm 60$ W/mK. 

\begin{figure}[hbt]
\begin{center}
\includegraphics[width=\columnwidth]{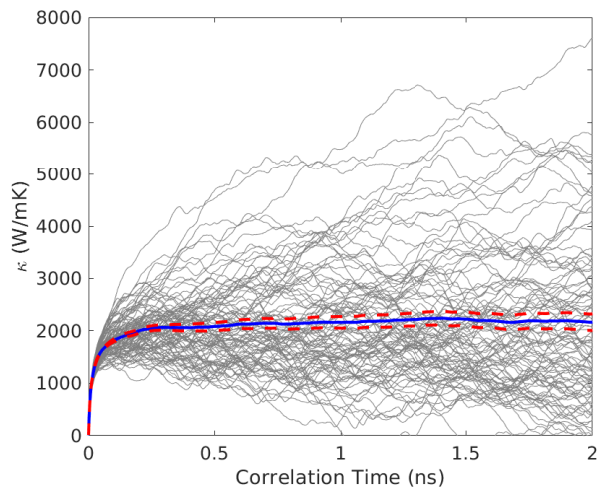}
\caption{Running thermal conductivity $\kappa(t)$ from the EMD method for $(10,10)$-CNT at 300 K as a function of the correlation time $t$. The thick solid line represents the average of the 100 thin solid lines (corresponding to 100 independent runs; each with a different set of initial velocities). The dashed lines represent the running statistical error bounds. }
\label{figure:cnt}
\end{center}
\end{figure}

To validate our HNEMD results for the $(10, 10)$-CNT, we performed EMD simulations for the same system. We performed $100$ independent runs, each with $20$ ns of \textit{production time}. All the other simulation parameters are the same as those for the HNEMD method. Figure \ref{figure:cnt} shows the running thermal conductivity $\kappa(t)$ as a function of the \textit{correlation time}. The averaged $\kappa(t)$ (thick solid line) converges well in the range of [1 ns, 2 ns] and we thus calculated 100 mean values in this range, from which we get a mean value and a standard statistical error (i.e., the standard deviation divided by the square root of the number of independent runs):  $\kappa=2200\pm 130$ W/mK. This is consistent with our HNEMD value.

In Refs. \cite{berber2000prl,lukes2006jht}, the driving forces were chosen to be in the range of $F_{\rm e}=500-4000$ $\mu$m$^{-1}$, which are several orders of magnitude larger than the threshold value above which linear response breaks down. Using these unphysically large driving forces, the authors \cite{berber2000prl,lukes2006jht} found that $\kappa(t)$ of $(10,10)$-CNT converges to about 100 W/mK within a couple of ps and the converged value increases with decreasing driving force. All these results deviate significantly from our results obtained in the linear response regime.

\subsection{Quantitative analysis of the computational efficiency and statistical errors}

From our benchmark in terms of silicon, graphene, and CNT, it is clear that the HNEMD method is much more efficient than the EMD method. The superior efficiency of this method over the EMD and NEMD ones has also been recently demonstrated in several studies \cite{xu2018msmse,dong2018pccp,xu2019prb}. To make it more quantitative, we take the case of CNT as an example and compare the relative computational efficiency of the HNEMD and EMD methods by examining the statistical errors in more detail. 

To this end, we first determine the statistically independent data in each method. For the EMD method, we already have 100 statistically independent $\kappa$ values from the 100 independent runs. For the HNEMD method, because we directly measure the nonequilibrium heat current, we can divide the whole production time into small blocks and take the $\kappa$ values calculated within different time blocks as independent values. Here, we consider a single HNEMD simulation with a production run of $20$ ns (the same as that for a single EMD simulation) and divide the total production time into 100 blocks, calculating 100 independent $\kappa$ values. 

\begin{figure}[hbt]
\begin{center}
\includegraphics[width=\columnwidth]{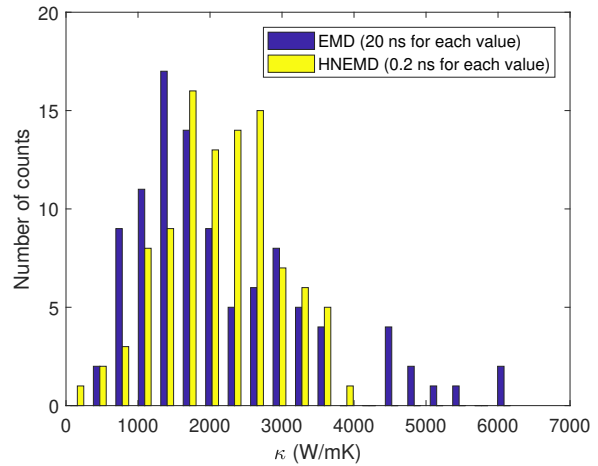}
\caption{Distribution (number of counts) of the $\kappa$ values from the EMD and HNEMD methods. Each $\kappa$ value in the EMD method is calculated based on a production time of $20$ ns, while each $\kappa$ value in the HNEMD method is calculated from a production time of only $0.2$ ns. }
\label{figure:cost}
\end{center}
\end{figure}

Figure \ref{figure:cost} shows that the distributions of both sets of $\kappa$ values have comparable variances and therefore comparable statistical errors. Because the total production time in the EMD method is 100 times as long as that in the HNEMD method, we see that the HNEMD method is about two orders of magnitude more efficient than the EMD method. The reason for the superior efficiency of the HNEMD method over the EMD method is related to the fact that in the EMD method, one measures the heat current autocorrelation function (HCACF), while in the HNEMD method, one directly measures the heat current. Because the noise-to-signal ratio in the decaying HCACF increases with increasing correlation time, the integrated running thermal conductivity has large variations in the limit of long correlation time where the averaged thermal conductivity converges. In contrast, the heat current measured in the HNEMD simulation has a constant noise-to-signal ratio (because it is not decaying) and the running average of the heat current converges quickly. 

Because we usually only need to perform a few independent simulations (or even a single one with relatively long production time) for a given system when using the HNEMD method, it is more practical to use the above time-block method to define the statistical error. That is, we first divide the total production time into a number of time blocks and obtain mean $\kappa$ values for all the time blocks. Then we calculate the statistical error as the standard deviation divided by the square root of the number of time blocks. We use this method to estimate the statistical errors as reported above.

\section{Spectral decomposition}

\subsection{Formalism}

An additional advantage of the HNEMD method is that the nonequilibrium heat current can be spectrally decomposed, similar to the case of the NEMD method \cite{fan2017prb,saaskilahti2015prb,zhou2015prb_b}. To this end, we define the following steady-state time correlation function:
\begin{align}
\label{equation:K}
\vect{K}(t) 
= \sum_i \sum_{j\neq i} \vect{r}_{ij}(0) \left\langle \left(\frac{\partial U_j}{\partial \vect{r}_{ji}}(0) \cdot\frac{\vect{p}_i(t)}{m_i} \right) \right\rangle_{\rm ne}, 
\end{align}
which reduces to the nonequilibrium heat current (the potential part) when $t=0$. Then one can define the following Fourier transforms:
\begin{equation}
\label{equation:K_omega_from_time}
\widetilde{\vect{K}}(\omega) = \int_{-\infty}^{\infty} dt e^{i\omega t} \vect{K}(t); ~
\vect{K}(t)= \int_{-\infty}^{\infty} \frac{d\omega}{2\pi} e^{-i\omega t} \widetilde{\vect{K}}(\omega).
\end{equation}
Setting $t=0$ in the second equation above yields the following spectral heat current (SHC):
\begin{equation}
\vect{J}_{\rm q}(\omega)= 2\widetilde{\vect{K}}(\omega); 
~
\langle\vect{J}_{\rm q}\rangle_{\rm ne} = \int_{0}^{\infty} \frac{d\omega}{2\pi} \vect{J}_{\rm q}(\omega).
\end{equation}
From the SHC, one can naturally get the spectral thermal conductivity (the vector $\vect{\kappa}$ denotes the diagonal part of the conductivity tensor):
\begin{equation}
\vect{\kappa}(\omega)= \frac{2\widetilde{\vect{K}}(\omega)}{TVF_{\rm e}}; 
~
\vect{\kappa} = \int_{0}^{\infty} \frac{d\omega}{2\pi} \vect{\kappa}(\omega).
\end{equation}
To our knowledge this is the only spectral decomposition method that works in the diffusive regime and does not require lattice dynamics calculations, which makes it applicable to spatially complex structures. The spectral decomposition also allows one to include quantum statistical corrections when appropriate \cite{lv2016njp,fan2017nl}. Below, we demonstrate the usefulness of this method by applying it to graphene and the $(10,10)$-CNT.

\subsection{Applications to graphene}

\begin{figure}[hbt]
\begin{center}
\includegraphics[width=\columnwidth]{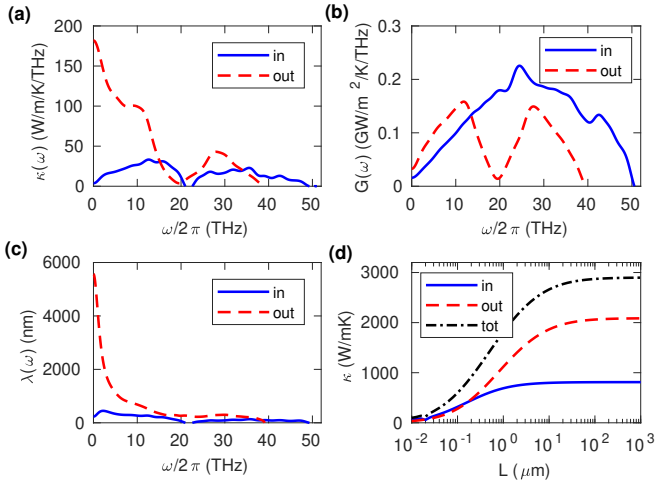}
\caption{The spectral thermal conductivity $\kappa(\omega)$ (a), the spectral ballistic conductance $G(\omega)$ (b), and the phonon mean free path $\lambda(\omega)$ (c) of graphene at 300 K as a function of the phonon frequency $\omega/2\pi$. (d) The length dependent thermal conductivity $\kappa(L)$. }
\label{figure:gra_spectral}
\end{center}
\end{figure}

Figure \ref{figure:gra_spectral}(a) shows the calculated spectral thermal conductivity $\kappa(\omega)$ of graphene at 300 K, for both the in-plane and the out-of-plane (flexural) phonons. It is clearly seen that the thermal conductivity of graphene is dominated by the flexural modes \cite{lindsay2010prb2,fan2017prb}. Moreover, we can also calculate the ballistic conductance $G(\omega)$  using the NEMD-based SHC \cite{fan2017prb,saaskilahti2015prb,zhou2015prb_b} and then obtain the spectral phonon MFP $\lambda(\omega)$ from (see Appendix \ref{section:spectral-derivation} for a derivation of Eqs. (\ref{equation:lambda_omega}) and (\ref{equation:kappa_omega}))
\begin{equation}
\label{equation:lambda_omega}
    \lambda(\omega)=\frac{\kappa(\omega)}{G(\omega)}.
\end{equation}
The calculated $G(\omega)$ and $\lambda(\omega)$ are shown in Figs. \ref{figure:gra_spectral}(b) and \ref{figure:gra_spectral}(c), respectively. From the spectral decomposition
\begin{equation}
\kappa(L)=\int_0^{\infty} \frac{d\omega}{2\pi} \kappa(\omega,L)
\end{equation}
and the ballistic-to-diffusive relation
\begin{equation}
\label{equation:kappa_omega}
\frac{1}{\kappa(\omega,L)}=\frac{1}{\kappa(\omega)}
\left(1+\frac{\lambda(\omega)} {L} \right),
\end{equation}
we can obtain the length dependent thermal conductivity $\kappa(L)$, as shown in Fig. \ref{figure:gra_spectral}(d). The large phonon MFP (a few microns) for the acoustic flexural (ZA) modes is responsible for the slow length convergence of the thermal conductivity of graphene as observed experimentally \cite{xu2014nc}. 

\begin{figure}[hbt]
\begin{center}
\includegraphics[width=\columnwidth]{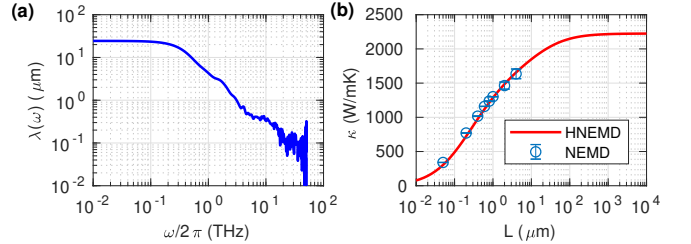}
\caption{(a) The phonon MFP $\lambda(\omega)$ of the $(10,10)$-CNT at 300 K and (b) the length dependent thermal conductivity $\kappa(L)$. The NEMD data are from Ref. \cite{saaskilahti2015prb}.}
\label{figure:cnt_spectral}
\end{center}
\end{figure}

\subsection{Applications to carbon nanotubes}

Last, we employ the HNEMD-based spectral decomposition method to examine the long-standing dispute over the thermal conductivity convergence vs. divergence in CNTs \cite{mingo2005nl,donadio2007prl,savin2009prb,lindsay2009prb,saaskilahti2015prb,liu2017nanoscale,lee2017prl,li2017prl,lee_arxiv,li_arxiv}. Figure \ref{figure:cnt_spectral}(a) shows that the phonon MFP scales as $\lambda(\omega)\sim \omega^{-1}$ for $\omega/2\pi > 0.25$ THz but saturates to about $\lambda_{\rm max}=25$ $\mu$m in the $\omega\to 0$ limit. This large value of $\lambda_{\rm max}$ dictates the small threshold value of $F_{\rm e} \leq 0.04$ in accordance with the criteria $F_{\rm e}\lambda_{\rm max}  \lesssim 1$.
The conductivity $\kappa(L)$ only fully converges when $L\approx 1$ mm, as shown in Fig. \ref{figure:cnt_spectral}(b). Our $\kappa(L)$ values agree well with the NEMD data with $L\leq 4$ $\mu$m by  S\"a\"askilahti \textit{et al.} \cite{saaskilahti2015prb}. However, because the large simulation cell sizes required in the NEMD method, it is computationally prohibitive to use this method to reach longer systems and they failed to obtain the $\lambda(\omega)$ values with $\omega/2\pi<0.25$ THz and could not resolve the issue of thermal conductivity convergence/divergence in CNTs. In contrast, our HNEMD-based spectral decomposition method can easily reach the diffusive regime and our results clearly demonstrate that $\kappa(L)$ in CNTs is upper bounded.

\section{Summary and Conclusions}

In summary, we have extended the HNEMD method for lattice thermal conductivity calculations with general many-body potentials. The method is about two orders of magnitude more efficient than EMD. A method for obtaining the spectral thermal conductivity and phonon mean free path is also developed based on HNEMD. This method works in the diffusive regime and does not require lattice dynamics calculations, making it suitable for studying spatially complex structures. Applying the spectral decomposition method, we find that the thermal conductivities of graphene ad CNTs converge with increasing length, but very slowly.

\begin{acknowledgments}
This work was supported by the NSFC (11404033) and the  Academy  of  Finland  Centre  of  Excellence program QTF (Project No. 312298). We acknowledge the computational resources provided by Aalto Science-IT project and Finland’s IT Center for Science (CSC).
\end{acknowledgments}

\appendix

\section{Thermal conductivity decomposition\label{section:decomposition}}

The heat current Eq. (\ref{equation:J_many-body}) is an extensive quantity consisting of many individual contributions. Therefore, it can be decomposed as
\begin{equation}
\vect{J}_{\rm q} = \vect{J}^{(1)}_{\rm q} + \vect{J}^{(2)}_{\rm q} + \cdots.
\end{equation}
This decomposition can be in terms of either real space \cite{matsubara2017jcp} or reciprocal space \cite{lv2016njp}. According to Eq. (\ref{equation:kappa}), this directly leads to a decomposition of the thermal conductivity:
\begin{equation}
\kappa(t) = \kappa^{(1)}(t) + \kappa^{(2)}(t) + \cdots = \frac{\langle J^{(1)}_{\rm q} \rangle_{\rm ne}}{TVF_{\rm e}} + \frac{\langle J^{(2)}_{\rm q} \rangle_{\rm ne}}{TVF_{\rm e}} + \cdots.
\end{equation}
According to an identity derived in the main text
\begin{equation}
\langle \vect{J}_{\rm q}(t)\rangle_{\rm ne}
=
\left(
\frac{1}{k_{\rm B}T}\int_0^tdt'\langle \vect{J}_{\rm q}(t')\otimes \vect{J}_{\rm q}(0)\rangle
\right)
\cdot \vect{F}_{\rm e},
\end{equation}
we can write an expression for the thermal conductivity decomposition in the EMD method:
\begin{equation}
\label{equation:k_i_emd}
\kappa^{(i)}(t) = \frac{1}{k_{\rm B}T^2V}\int_0^tdt'\langle J_{\rm q}^{(i)}(t') J_{\rm q}(0)\rangle.
\end{equation}
This is the same result proved in Ref. \cite{matsubara2017jcp}.

In the main text, we have considered an in-out decomposition of the potential part of the heat current:
\begin{equation}
\vect{J}_{\rm q}^{\rm in} =
\sum_i \sum_{j\neq i} \vect{r}_{ij} \left(\frac{\partial U_j}{\partial x_{ji}}  \frac{p_{ix}}{m_i} +\frac{\partial U_j}{\partial y_{ji}}  \frac{p_{iy}}{m_i} \right);
\end{equation}
\begin{equation}
\vect{J}_{\rm q}^{\rm out} =
\sum_i \sum_{j\neq i} \vect{r}_{ij} \left(\frac{\partial U_j}{\partial z_{ji}}  \frac{p_{iz}}{m_i} \right).
\end{equation}
The in-plane heat current only involves in-plane phonons and the out-of-plane heat current only involves out-of-plane (flexural) phonons. This heat current decomposition leads to a decomposition of the thermal conductivity
\begin{equation}
\kappa(t) = \kappa^{\rm in}(t) + \kappa^{\rm out}(t) = \frac{\langle J^{\rm in}_{\rm q} \rangle_{\rm ne}}{TVF_{\rm e}} + \frac{\langle J^{\rm out}_{\rm q} \rangle_{\rm ne}}{TVF_{\rm e}}.
\end{equation}
According to Eq. (\ref{equation:k_i_emd}), we see that the ``crossterm'' defined in Ref. \cite{fan2017prb} should be evenly attributed to the in-plane and out-of-plane parts defined there.

\section{Spectral conductivity and spectral mean free path\label{section:spectral-derivation}} 

In macroscopic transport, thermal conductance $G$ (per unit area) is related to thermal conductivity $\kappa$ by:
\begin{equation}
G = \frac{\kappa}{L},
\label{equation:K_and_kappa_no_L}
\end{equation}
where $L$ is the system length. Usually, the thermal conductivity is an intrinsic property of a material. At the nanoscale, however, the conventional concept of the conductivity can become invalid \cite{datta1995} and the conductivity as defined in  Eq. (\ref{equation:K_and_kappa_no_L}) is length dependent, $\kappa=\kappa(L)$. We therefore write
\begin{equation}
G(L) = \frac{\kappa(L)}{L}.
\label{equation:K_and_kappa}
\end{equation}
This length dependence can be captured by noticing that there is a resistance $1/G_0$ even in the ballistic limit due to the finite number of conducting channels. For a system of length $L$, the total resistance comes from the ballistic resistance and a length dependent resistance \cite{datta1995}: 
\begin{equation}
\frac{1}{G(L)} = \frac{1}{G_0} +  \frac{L}{\kappa_{\rm diff}}.
\label{equation:K_and_kappa_diff}
\end{equation}
Here, $\kappa_{\rm diff}$ is the conductivity in the diffusive limit, i.e., $\kappa_{\rm diff}=\kappa(L\to \infty)$. By comparing Eq. (\ref{equation:K_and_kappa}) and Eq. (\ref{equation:K_and_kappa_diff}), we have the following relation between the length dependent thermal conductivity $\kappa(L)$ and the length independent diffusive thermal conductivity $\kappa_{\rm diff}$:
\begin{equation}
 \frac{L}{\kappa(L)} = \frac{1}{G_0} +  \frac{L}{\kappa_{\rm diff}},
\label{equation:kappa_and_kappa_diff}
\end{equation}
or equivalently,
\begin{equation}
\frac{1}{\kappa(L)} = \frac{1}{\kappa_{\rm diff}} 
\left( 
1+ \frac{\kappa_{\rm diff} /G_0}{L}
\right).
\end{equation}
Comparing this with the standard length scaling formula of conductivity,
\begin{equation}
\frac{1}{\kappa(L)} = \frac{1}{\kappa_{\rm diff}} 
\left( 
1+ \frac{\lambda}{L}
\right),
\label{equation:ballistic_to_diffusive_lambda}
\end{equation}
we see that the ratio between the diffusive conductivity and the ballistic conductance defines a phonon mean free path (MFP) in an infinite system:
\begin{equation}
\lambda = \frac{\kappa_{\rm diff}}{ G_0}.
\label{equation:mfp_from_G}
\end{equation}
The length scaling formula Eq. (\ref{equation:ballistic_to_diffusive_lambda}) can be derived from Matthiessen's rule,
\begin{equation}
\frac{1}{\lambda(L)} = \frac{1}{\lambda} + \frac{1}{L},
\end{equation}
and the relation
\begin{equation}
\frac{\kappa_{\rm diff}}{\kappa(L)} = \frac{\lambda}{\lambda(L)}.
\end{equation}
Here, $\lambda(L)$ is the effective MFP in a finite system of length $L$ whose conductivity is $\kappa(L)$.

The above discussion is simplified in the sense that no frequency dependence of the thermal transport has been taken into account. Different frequencies usually have different MFPs and diffusive conductivities. In general, both the conductivity and the MFP are frequency dependent and we can generalize Eqs. (\ref{equation:ballistic_to_diffusive_lambda}) and (\ref{equation:mfp_from_G}) to
\begin{equation}
\frac{1}{\kappa(\omega,L)} = \frac{1}{\kappa_{\rm diff}(\omega)} 
\left( 1 + \frac{\lambda(\omega)}{L} \right);
\label{equation:ballistic_to_diffusive_lambda_omega}
\end{equation}
\begin{equation}
\lambda(\omega) \equiv \frac{\kappa_{\rm diff}(\omega)}{ G_0(\omega)}.
\label{equation:mfp_from_G_omega}
\end{equation}
Note that $\kappa_{\rm diff}(\omega)$ and $G_0(\omega)$ were respectively written as $\kappa(\omega)$ and $G(\omega)$ in the main text. We thus have derived  Eqs. (\ref{equation:lambda_omega}) and (\ref{equation:kappa_omega}).

\end{document}